\documentclass[12pt,preprint]{aastex}

\shorttitle{ISM Faraday-Rotation Structure}
\shortauthors{Uyan{\i}ker \& Landecker}

\begin{document}

\title{A Highly Ordered Faraday-Rotation Structure in the Interstellar
Medium}

\author{B. Uyan{\i}ker and T.L. Landecker}
\affil{National Research Council, Herzberg Institute of Astrophysics,
Dominion Radio Astrophysical Observatory, Penticton, B.C., Canada V2A 6K3}
\email{bulent.uyaniker@nrc.ca;tom.landecker@nrc.ca}

\begin{abstract}

We describe a Faraday-rotation structure in the Interstellar Medium
detected through polarimetric imaging at 1420 MHz from the Canadian
Galactic Plane Survey (CGPS).  The structure, at
${l}={91\fdg8},{b}={-2\fdg5}$, has an extent of ${\sim}2^{\circ}$,
within which polarization angle varies smoothly over a range of
${\sim}100^{\circ}$. Polarized intensity also varies smoothly, showing
a central peak within an outer shell.  This region is in sharp
contrast to its surroundings, where low-level chaotic polarization
structure occurs on arcminute scales.  The Faraday-rotation structure
has no counterpart in radio total intensity, and is unrelated to known
objects along the line of sight, which include a Lynds Bright Nebula,
LBN~416, and the star cluster M39 (NGC7092).  It is interpreted as a
smooth enhancement of electron density.  The absence of a counterpart,
either in optical emission or in total intensity, establishes a lower
limit to its distance.  An upper limit is determined by the strong
beam depolarization in this direction. At a probable distance of
$350{\pm}50$~pc, the size of the object is 10 pc, the enhancement of
electron density is 1.7 cm$^{-3}$, and the mass of ionized gas is 23
M$_{\odot}$. It has a very smooth internal magnetic field of strength
3 $\mu$G, slightly enhanced above the ambient field. G91.8$-$2.5 is
the second such object to be discovered in the CGPS, and it seems
likely that such structures are common in the Magneto-Ionic Medium.

\end{abstract}


\section{Introduction}

Observation of the radio sky between 300 MHz and 3 GHz has revealed an
extensive polarized emission that has no counterpart in total
intensity 
\citep{junk87,wier93,dunc97,dunc99,gray98,gray99,uyan98,uyan99,have00,gaen01}, 
supplementing, and often dominating, polarized features associated
with identifiable sources such as Galactic supernova remnants (SNRs)
and compact non-thermal sources.  These unidentifiable polarized
features are detected as changes in the polarization angle of the
received signal; they are generally attributed to rapid changes in
Faraday rotation through a foreground Magneto-Ionic Medium (MIM)
acting on polarized emission from a smooth Galactic synchrotron
background.

Faraday rotation produces a change in polarization angle of
\begin{equation} {{\Delta}{\theta}} =
{0.81{\lambda^2}{\int}{n_e}{B_{\|}}{dL}} =
{{\lambda^2}RM}~{\rm{radians}} \end{equation} where $n_e$ (cm$^{-3}$)
is the electron density, $B_{\|}$ ($\mu$G) is the component of the
magnetic field parallel to the line of sight, $L$ (pc) is the path
length, and $\lambda$ (m) is the wavelength. RM is the Rotation
Measure. Changes in Faraday rotation imply that the intervening
``Faraday screen'' has structure in the intensity or direction of the
magnetic field or in the density of electrons or some combination.
Polarization observations therefore allow us to probe physical
conditions in the MIM.

In most of the regions imaged to date, the structure is chaotic, often
with features on arcminute scales. In this paper we investigate a
region of extent $2^{\circ}$ where polarization angle varies smoothly
over a range ${\Delta\theta}={100^{\circ}}$, and structure in Stokes
$Q$ and $U$ is smooth with no counterpart in total intensity. It
resembles the elliptical Faraday-rotation structure G137.6+1.1
described by \citet{gray98}, a region $1^{\circ} \times 2^{\circ}$
where polarization angle changes smoothly by ${200^{\circ}}$. 
G137.6+1.1 was interpreted as a region of smooth enhancement of
electron density, with a smooth magnetic field structure. The
existence of two such objects suggests that they may be common in the
Interstellar Medium (ISM).

This paper uses some results from a broader discussion of CGPS
polarization observations of the area ${82^{\circ}}<{l}<{95^{\circ}}$,
${-3.6^{\circ}}<{b}<{5.6^{\circ}}$ (Uyan{\i}ker et al. in
preparation), which we refer to as Paper 2.

\section{Observations and Data Processing}

The observations are part of the Canadian Galactic Plane Survey
\citep[CGPS]{tayl02} made with the Synthesis Telescope at the
Dominion Radio Astrophysical Observatory \citep{land00}.
Standard data reduction algorithms are described in \citet{will99}
and \citet{tayl02}.  The survey covers
${75^{\circ}}<{l}<{145^{\circ}}$, ${-3.6^{\circ}}<{b}<{5.6^{\circ}}$.
Within this region, images are obtained on a hexagonal grid with field
centers spaced by $117\arcmin$.  Angular resolution is $49\arcsec \times
49\arcsec~{\rm{cosec}}({\delta})$.

The telescope receives both hands of circular polarization in four
bands near 1420 MHz, each 7.5 MHz wide, set above and below the
\ion{H}{1} frequency by offsets of $\pm 6.25$ MHz and $\pm 13.75$ MHz.
For the observations presented here the center frequency was 1420.64
MHz.  Visibilities are processed into images of Stokes parameters $I$,
$Q$ and $U$ in each of the four bands using standard
aperture-synthesis techniques.  $Q$ and $U$ images are corrected for
instrumental polarization across the field of view, and are assembled
into mosaics. Images presented here are averages from all four bands.

The telescope samples baselines from 12.9 to 617 m with an increment
of 4.286 m, and all structure from the resolution limit of ${\sim}1\arcmin$
up to ${\sim}45\arcmin$ is represented in the images from the Synthesis
Telescope. Information from single-antenna data has been incorporated
into the $I$ data (as is standard practice for the CGPS), but no
attempt has been made to recover information corresponding to larger
polarization structure, and single-antenna data have not been added to
the $Q$ and $U$ images discussed in this paper.

Polarized intensity, $PI$, nominally ${(Q^2 + U^2)}^{1/2}$, is
calculated as \\  ${PI}={(Q^2 + U^2 - (1.2\sigma)^2)^{1/2}}$, where
${\sigma} = 30$ mK T$_{\rm B}$ is the rms noise in the mosaiced $Q$
and $U$ images, in order to provide a first-order correction for noise
bias. The polarization angle in each of the four bands is calculated
as
${{\theta}_{\lambda}}=\frac{1}{2}\thinspace{\rm{arctan}}\thinspace{(U/Q)}$.
RM can be calculated from the four values of ${\theta}_{\lambda}$ when
the signal-to-noise ratio is adequate.

\section{Results}

Figure~\ref{ipimap} shows $I$ and $PI$ images, and Figure~\ref{qumap}
shows $Q$ and $U$ images from the CGPS database in a region
${88^{\circ}}<{l}<{95^{\circ}}$, ${-3\fdg6}<{b}<{1^{\circ}}$.

Objects visible in the $I$ image include the SNRs CTB~104A (at
${l}={93\fdg7}, {b}={-0\fdg2}$) and 3C434.1 (at ${l}={94\fdg0},
{b}={1\fdg0}$, partly seen at the upper edge of the image). Strong
polarized emission from CTB~104A is evident in the $PI$, $Q$, and 
$U$ images, but no polarized emission corresponding to 3C434.1 has
been detected. The polarized emission from CTB~104A is discussed by
\citet{uyan02}. An \ion{H}{2} region, S~124, is seen at
${l}={94\fdg5}, ~{b}={-1\fdg5}$, consisting of a bright, compact
object embedded in extended emission. Diffuse, low-brightness emission
fills the $I$ image north of ${b}\approx{-0\fdg5}$.

Within the area shown in Figure~\ref{qumap}, at ${l}={91\fdg75}$,
${b}={-2\fdg45}$, is an elliptical region of extent nearly $2^{\circ}$
where the structure in $Q$ and $U$ changes very smoothly, indicating
that polarization angle changes slowly and systematically. No
counterpart in total intensity is detectable in
Figure~\ref{ipimap}. This unusual feature is the subject of this
paper. To give the object a name that reflects its appearance, we
refer to it as the Polarized Lens (PL), but this is not intended to
suggest any particular interpretation.  The contrast between the PL
and the SNR CTB104A is very striking. The SNR shows very strong $PI$
and chaotic, small-scale structure in $Q$ and $U$ while the PL shows
strong, smoothly structured $Q$ and $U$ emission but very low $PI$.
To the west and north-west of the PL there are regions of small-scale
chaotic structure, again with no counterpart in total intensity, whose
appearance is characteristic of the polarized emission in many
directions in the Galactic plane (see references cited in the
Introduction).

The area shown in Figures~\ref{ipimap} and~\ref{qumap} is at the
southern edge of the CGPS region, and the data become noisy at the
extreme edge. However, we are confident that the PL does not extend
south beyond the limit of these images.  An overlapping region
was observed by \citet{higg01}; those observations are consistent with
our data where they overlap, and extend well to the south of the field
presented here, showing very low levels of $Q$ and $U$.

Figures~\ref{tp} displays total intensity, $I$, scaled to show lower
intensities than are seen in Figure~\ref{ipimap}. Extended
emission is seen in the field; most of it is probably thermal in
origin. Some of this emission is related to known objects, also
indicated in Figure~\ref{tp}; these are discussed in detail below.
However, there is no extended emission which resembles the PL in form,
and we consider that the PL is a region where the polarized ``object''
has no counterpart in $I$; it is one of the Faraday objects described
in the Introduction.

Figure~\ref{tp} also shows polarized intensity, $PI$, and polarization
angle in a region centered on the PL. Polarization vectors are
superimposed on the $PI$ image. Because the polarized structure is
very smooth, the $PI$ and angle images have been derived from data
convolved to $5\arcmin$ resolution.  For reference, two ellipses are
superimposed on the images of Figure~\ref{tp}, both centered at
${l}={91\fdg75}$, ${b}={-2\fdg45}$.  The larger ellipse, $90\arcmin
\times 108\arcmin$, marks the approximate limit of the smooth
structure. The smaller ellipse, $72\arcmin \times 42\arcmin$,
approximately follows the inner contours in the polarization angle
image of Figure~\ref{tp}.

The emission here is quite strongly polarized, but very smooth. The
overall extent of the polarization feature, defined as the region in
which polarization angle varies smoothly, is about $100\arcmin$. Within this
region the polarized intensity, $PI$, rises to $\sim$0.08~K, while in
the immediate surroundings $PI$ is mostly below the sensitivity limit
of the telescope. 

Figure~\ref{prof} shows the variation of polarization angle along the
major axis of the larger ellipse. The change of angle through the PL
is $\sim{100}^{\circ}$, corresponding to a RM change of 40
rad~m$^{-2}$. The signal-to-noise ratio is inadequate to permit the
derivation of an image of RM.

\section{Possible Relationship to Other Objects in the Field}

In seeking to interpret the PL we have investigated possible
associations with a variety of objects in this direction. There are
many objects along this line of sight, which passes along the local
spiral arm close to the Galactic plane; some are identified in Figure~\ref{tp}.

\subsection{Possible supernova remnant}

\citet{kass88a} reported a source at
$(l,b)=(92.0^{\circ},-2.6^{\circ})$ with an extent of $44\arcmin \times
35\arcmin$, detected in the Clark Lake survey at 30.9 MHz (resolution
$15\arcmin$); the reported flux density at 30.9 MHz was 23 Jy. Kassim
originally identified this with 4C 47.56 at
$(l,b)=(92\fdg23,-2\fdg97)$.  However, in a subsequent
paper he suggested the Clark-Lake source might be an
SNR \citep{kass88b}.

We note that other low-frequency surveys do not indicate the presence
of an extended low-frequency source corresponding to the PL. No source
is detected in the 22.25 MHz DRAO survey \citep{roge99}, with
resolution $1\fdg1 \times 1\fdg7$ at this sky position. The
34.5 MHz Gauribidanur survey of \citet[resolution $26\arcmin \times 40\arcmin$]{dwar90}
shows a depression at this location.
\citet{gorh90} does not confirm G92.0$-$2.6 as an SNR from examination
of radio continuum surveys at 1420 and 2695 MHz. However, Gorham notes
a possibly related infrared feature in the IRAS 60$~\mu$m data. Our
examination of the CGPS infrared database \citep{cao97} has
revealed a depression in the infrared emission coinciding roughly with
the PL.

Nevertheless, the appearance of the PL in polarized intensity does
bear some resemblance to an SNR (see Figure~\ref{tp}). There is a
bright outer ridge of emission, and a central feature, defining a
morphology reminiscent of a composite SNR. The outer ridge shows
strong bilateral symmetry about the major axis of the smaller ellipse.
The integrated polarized intensity at 1420 MHz is $\sim$0.9 Jy, which,
assuming $<$50\% fractional polarization, implies a total flux density
of $>$2 Jy. With a spectral index of ${\alpha}={0.5}$
(${S}\propto{{\nu}^{-\alpha}}$), the flux density would be $>$14 Jy at
30.9 MHz, plausibly consistent with the flux density of 23 Jy reported
by Kassim.  

However, the smooth polarization structure does not resemble that of a
typical SNR. Those SNRs whose polarized emission {\it{is}} seen in
this vicinity show generally chaotic structure on arcminute scales (as
demonstrated by CTB~104A -- see Figure~\ref{qumap}).  Young SNRs have
radial magnetic field, producing polarization vectors which follow the
circumference. Old SNRs have circumferential field, and radial E
vectors. To produce the observed pattern of polarization vectors 
would require a very special distribution of Faraday rotation,
and we consider this an unlikely circumstance.

4C 47.56 (at $(l,b)=(92.0^{\circ},-2.6^{\circ})$) lies very close to
the eastern boundary of the larger ellipse (see Figure~\ref{tp}). Its
location relative to the structure seen in polarized intensity
suggests it could be related to the PL. However, its spectral index,
derived from published flux densities between 151 and 4850 MHz,
together with values derived from our own 1420 MHz data, is
${\alpha}={1.03{\pm 0.04}}$, typical of an extragalactic source. The
predicted flux density at 30.9 MHz is 24.8 Jy, entirely consistent
with the value obtained by \citet{kass88a}.  There is thus no excess
flux density at 30.9 MHz which might require the hypothesized SNR.

\subsection{\ion{H}{1} feature G92$-$4+50}

\citet{higg01} report an unusual \ion{H}{1} structure at
$(l,b){\simeq}(92^{\circ},-4^{\circ})$, with a velocity of
${v_{LSR}}={50}$ km s$^{-1}$ which partially overlaps the PL in
position. However, we can find no morphological resemblance between
this \ion{H}{1} structure, or any other \ion{H}{1} structure in the
field, and the PL.

\subsection{\ion{H}{2} Region LBN~416}

\citet{lynd65} lists a filamentary emission nebula, LBN~416 (indicated
in Figure~\ref{tp}) crossing the position of the PL. The position and
size, determined by Lynds from the Palomar Sky Survey, are
$(l,b)=(91\fdg66,-2\fdg54)$ and $80\arcmin \times 10\arcmin$. The object
is seen clearly in total intensity in our observations. CGPS data at
408 MHz confirm that the emission from LBN~416 has a thermal spectrum,
and we therefore consider it to be an \ion{H}{2} region.

The peak brightness temperature of LBN~416 is 0.2K, and the integrated
flux density at 1420 MHz is roughly 0.7 Jy, with a measured extent of
$90\arcmin \times 15\arcmin$. We have modeled the \ion{H}{2} region as
a cylinder of ionized gas (using a program written by L.A. Higgs). We
obtain reasonable fits to the observed parameters with electron
densities of 8 and 2 cm$^{-3}$ for assumed distances of 0.25 and 2.5
kpc respectively, with corresponding variation for intermediate
distances. We assume that the filament is cylindrical, so that the
line-of-sight depth through it is equal to its transverse
extent. Further assuming a magnetic field of 2 $\mu$G, the Faraday
rotation through the filament would be $36^{\circ}$ at a distance of
0.25 kpc, and $120^{\circ}$ at a distance of 2.5 kpc.  If LBN~416 were
in front of the PL, such rotations would easily be detected in our
data.  The fact that they are not, implies that LBN 416 is
{\it{behind}} the PL.  In fact, the absence of any polarization effect
attributable to LBN 416 probably implies that it is more distant than
2 kpc (see Section 5.3 where we discuss the distance of detectable
polarization features).  We conclude that LBN 416 and the PL are
unrelated.

\subsection{Stars in the field}

An open star cluster, M39 (NGC7092), partly overlaps the PL; its
distance is 265 pc \citep{mcna77} and the PL could
conceivably be produced by effects of one or more of its stars. The
nominal cluster center is $(l,b)=(92\fdg2,-2\fdg5)$.  Of
7931 cataloged stars in the field, roughly 60 have a probability of
membership $>$90\% \citep{plat94}, and these are indicated in Figure~\ref{tp}.
The brightest 25 of these stars range in spectral type from B8 to
A0, and their combined ionizing flux might form a Str\"omgren sphere
of diameter a few pc, dependent on local conditions of ambient
density. This is too small to account for the PL.

There are two B stars within the boundary of the PL.  The B3 star,
No. 3721 in \citet{plat94}, lies at $(l,b)= (92\fdg12,-2\fdg51$).  The
Tycho catalog \citep{hog00} gives its parallactic distance as 300
pc. Although it is at a comparable distance to M39, \citet{plat94} did
not include this star in the membership list.  Its color index is
$(B-V)=-0.^{\rm m}24$. However, such a star can ionize a region of
order 1 pc in diameter in a region of ${n_e}={{1}~{\rm{cm}}^{-3}}$,
subtending an angle of only ${\sim}10\arcmin$ at the distance of the
star, much smaller than the PL.

The second is a B0 star \citep[LS $+47^{\circ}$ 47, No. 4071]{plat94}
close to the center of the PL, at $(l, b)= (92\fdg02, -2\fdg71$); this
star is probably not a member of M39.  The star has a color index
$(B-V)=-0.^{\rm m}13$. The absolute magnitude of a main sequence B0
star, on the other hand, is $M_V=-3.^{\rm m}4$. These parameters yield
a distance about 12 kpc for the star if it is on the main sequence. To
bring this star within 1 kpc, where it might conceivably be related
to the PL, it must have evolved to the point where it has spawned a
planetary nebula. No such nebula has been detected.

The offset between the PL and M39 makes an association between the two
objects unlikely, and the cluster has insufficient ionizing flux to
generate the PL. Of the two B stars in the field one has inadequate
ionizing flux to explain the PL and the other is too distant. We
cannot explain the PL on the basis of known stars in the field.

\section{Discussion}

\subsection{Is the PL an isolated structure?}

The images of polarized emission shown in Figures~\ref{qumap} and \ref{tp} suggest that
the PL is an isolated structure. However, the DRAO Synthesis Telescope
has limited sensitivity to structures larger than $45\arcmin$ for all Stokes
parameters.  It is important to consider whether our data are missing
any component that is vital to the correct interpretation of the PL.

There is no doubt that we have detected significant structure in $Q$
and $U$ from the PL. We have also observed an absence of $Q$ and $U$ in
the areas surrounding the PL, and this too is a significant result.
Nevertheless, the question remains: is there a large polarized object
in this direction, of which the Synthesis Telescope has seen only a
small part? The best answer to that question would be provided by
complementary single-antenna observations which could be appropriately
combined with the Synthesis Telescope data, but such data are not
available. The best we can do in answering that question is to turn to
the existing single-antenna data.

We have examined the data of \citet{brou76} for this region, observed
with a $36\arcmin$ beam at 1420 MHz. In the area
${90^{\circ}}<{l}<{94^{\circ}}$, ${-4^{\circ}}<{b}<{0^{\circ}}$ there
are 6 observations. Polarized intensity varies from 0.05 to 0.13 K,
except for one corner of the region where the value is 0.39 K (the
mean error of these observations is quoted as 0.06 K). The
polarization angle throughout the region is $145^{\circ} \pm
25^{\circ}$. Considering the errors and the severe undersampling,
these data support a picture of a region of roughly constant polarized
intensity and angle. The peak polarized intensity of the PL measured
in our data, ${\sim}0.08$~K, is comparable to the polarized intensity
of the ``background'' measured by Brouw and Spoelstra. Examination of
other data from the same paper at four lower frequencies (down to 408
MHz) also shows a low-level, fairly smooth structure.  The smooth
component is likely to be a foreground, since distant, and
consequently small, structure will average to a very low value in the
broad beam of the telescope used by Brouw and Spoelstra.

We therefore proceed on the assumption that our observations have
``seen'' all of the PL.  In this connection, we note that G137.6+1.1
was apparently detected in its entirety by the DRAO Synthesis
Telescope in the observations described by \citet{gray98}.
Observations with the Effelsberg 100-m Telescope at 1420 MHz
(W. Reich, private communication) confirm this, and show G137.6+1.1
superimposed on a uniform but intense polarized background.

\subsection{Is the PL an ionized or a magnetic feature?}

The morphology of the PL gives the impression of a confined region,
and in the following discussion we assume that the PL has a depth
along the line of sight equal to its observed extent on the sky,
approximately $29D$ pc, where $D$ is the distance in kpc. We consider
it unlikely that the PL is produced by a long cylindrical structure at
a fortuitous alignment.  We assume that the regular and random
components of the local magnetic field have values $2{~\mu}$G and
$4{~\mu}$G respectively \citep{beck01}, and that the regular component
is directed towards ${l}\approx{83^{\circ}}$ \citep{heil96}.

Through analysis of RMs of extragalactic point sources from the CGPS
database, \citet{brow01} have delineated a region of size
${\sim}2\fdg5$ centered on ${l}={92^{\circ}}$, ${b}={0^{\circ}}$
where the magnetic field reverses its direction, refining an earlier
detection by \citet{cleg92}. The cause of the field reversal has
not been identified.  The PL lies at the southern edge of this
magnetic anomaly and the question arises whether the two are
related. The SNR CTB~104A also lies on the edge of this region, and
\citet{uyan02} have shown that the magnetic field direction
in the vicinity of the SNR is opposite to that of the prevailing
magnetic field, and is aligned with the field in the anomaly. The SNR
and the anomaly are therefore at the same distance. \citet{uyan02} 
also show that the distance to CTB~104A is approximately 1.5
kpc.  We show below that the PL is much closer, at about 600 pc. The
PL is therefore unrelated to the magnetic anomaly.

Nevertheless, the possibility that the PL is a magnetic phenomenon
must be considered.  A change in Faraday rotation such as that
generated by the PL must be associated with an excess or deficit of
electrons, or with a change in the magnetic field intensity or
direction, or with some combination of the two. If the excess RM in
the PL is produced by a magnetic field fluctuation alone, the required
change is ${\sim}15 ~\mu$G for ${n_e}={{0.2}~{\rm{cm}}^{-3}}$,
considerably larger than the expected random field fluctuations, and
we consider it unlikely that the PL is solely a magnetic field
phenomenon. This leaves a deficit or excess of electrons.  The
required change in $n_e$ is of the order of 1~cm$^{-3}$, and a deficit
of this magnitude can only be accommodated if the surrounding electron
density is quite high. Further, a region where there is a deficit of
electrons relative to the surroundings must be filled with neutral gas
to maintain pressure balance, and neither atomic or molecular material
is in evidence (although the sensitivity of available observations
does not allow us to categorically exclude such an occurrence). We
proceed, therefore, on the assumption that the PL is the result of a
local enhancement of electron density. This enhancement is possibly
accompanied by a magnetic field structural change, probably also an
enhancement, but the magnetic field must be very smooth. In
considering G137.6+1.1, \citet{gray98} have presented very similar
arguments.

\subsection{General remarks on the distance to polarization features}

In this direction our line of sight is about $10^{\circ}$ from
the axis of the local spiral arm, which we judge to coincide with the
brightest emission from the Cygnus-X region at ${l}={80^{\circ}}$.
The line of sight is thus nearly parallel to the local field
direction, and Faraday rotation is strong. It is shown in Paper 2 that
the Faraday screen that generates the bulk of the polarized signal in
this region is within the local arm.  Polarized emission originating
at larger distances, either in distinct synchrotron sources such as
SNRs or in more distant Faraday screens, is rendered undetectable by
beam depolarization in the local Faraday screen. Depth depolarization
is also a strong effect. (Beam depolarization is also referred to as
differential Faraday rotation and depth depolarization as internal
Faraday dispersion; these effects are discussed by \citet{soko98}).
Paper 2 presents evidence that this distance, which is referred to
there as the {\it{polarization horizon}}, is about 2 kpc. Some of the
evidence is seen in Figure~\ref{ipimap}. Polarized emission is detected
from the SNR CTB~104A (G93.7$-$0.2), which is within the Local arm
\citep{uyan02}, but no polarized emission is detected from the SNR
3C434.1 (G94.0+1.0), which is a Perseus Arm object \citep{fost00}. For
more detailed arguments about the distance to the polarization
horizon, the reader is referred to Paper 2.

Our first conclusion is that the PL must be closer than about 2
kpc. However, in a direction where total beam depolarization occurs
over a distance of 2 kpc, a feature with the highly regular
polarization structure of the PL must be much closer.

\subsection{Considerations based on synchrotron emissivity}

The peak polarized intensity on the PL is about 0.08 K. Assuming that
the emission is 70\% polarized at source, the emitted intensity is at
least 0.11 K. This emission must be generated along a path
within the polarization horizon or it could not be detected. This
value is a lower limit: when the detectable polarized emission is
produced by Faraday rotation, the level of polarized intensity
detected is strongly dependent on the rotation in the foreground
screen.

The synchrotron emissivity at 408 MHz has been determined by \citet{beue85},
based on an analysis of a whole-sky survey at that
frequency.  These authors obtain a value equivalent to 11 K per kpc of
path at 408 MHz. When translated to 1420 MHz using a temperature
spectral index ${\beta}={2.8}$, the value is 0.26 K/kpc.
\citet{roge99} measured the local synchrotron emissivity at 22 MHz
by observing the brightness temperature against foreground \ion{H}{2}
regions, which absorb more distant emission. They observed two \ion{H}{2}
regions close to the PL, S~117 (G85.5$-$1.0) at a distance
of 800 pc, and S~131 (G99.3+3.7) at a distance of 860 pc. The
two values obtained were 21,800 K/kpc and 43,500 K/kpc. The average value
from eight \ion{H}{2} regions, 30,000 K/kpc, well represents local
conditions. Translating this value to 1420 MHz, again using ${\beta}={2.8}$,
gives 0.33 K/kpc.

Taking the mean of these two determinations of the local emissivity,
we can calculate that a path length of about 400 pc is required to
generate 0.11 K of synchrotron emission.  There must therefore be at
least 400 pc of path length within the polarization horizon, behind
the PL.  The maximum distance to the PL is therefore $\sim$1.6 kpc.
Again, this is not a severe constraint, and
the smooth structure implies a closer distance.

\subsection{Distance constraints based on the properties of ionized regions}

Constraints on the distance to the PL can be obtained from
consideration of the properties of emission from ionized gas, and
exploring the properties of a Faraday rotating region which can
produce the observed rotation angle, $\theta$, at 1420 MHz while
remaining undetectable in total intensity. We assume that the PL is a
region of enhanced electron density, but note that the following
arguments apply equally to a deficit of electrons.

The optical depth of a region of free-free emission is 
\begin{equation}
\tau = 8.235 \times 10^{-2} ~\left( { {T_e} \over {\rm K} } \right)^{-1.35}
                            ~\left( { {\nu} \over {\rm GHz} } \right)^{-2.1}
                            ~\left( { {E} \over {\rm pc ~cm^{-6}} } \right)
\end{equation}
where ${T_e}$ is the electron temperature and ${E}={n_e^2L}$ is the
emission measure \citep{rohl96}.  
The brightness temperature is related to $\tau$ by
\begin{equation}
{T_b} = {T_e} ~\left( 1 - {\rm e}^{-\tau} \right) 
\end{equation}

We assume that the PL has electron temperature $T_e$ = 7000 K, a value
typical of Galactic \ion{H}{2} regions, and we will consider initially
a magnetic field of $B_{\|} = 2~\mu$G. The angular extent of the
source is $\theta=100\arcmin$ and we assume a spherical ionized region
whose diameter is determined from its angular size. We then calculate
$n_e$ for various values of $\Delta\theta$, and derive the anticipated
brightness temperature and emission measure. The results of this
calculation are plotted in the top two panels of
Figure~\ref{em}. Since we do not detect the PL optically, or as a
feature in the $I$ image, we can constrain its distance.  The
theoretical continuum sensitivity of the DRAO Synthesis Telescope is
0.071~sin$\delta$~K~rms, where $\delta$ is the declination
\citep{land00}.  For the present observations the value achieved in
areas free of complex structure is $\sim$0.05 K, consistent with
expectation.  In the top panel of Figure~\ref{em} we show the
2$\sigma$ and 4$\sigma$ values.  


The observed value of ${\Delta\theta}={100^{\circ}}$ and the
non-detection of the PL in $I$
constrain the distance to be larger than about 400 pc.  On the other
hand, we have already established that it must be closer than
${\sim}1.6$ kpc. Furthermore, if it is close to that limit, we would
expect superimposed foreground Faraday rotation to detract from its
smooth appearance. We therefore choose a somewhat arbitrary upper
limit of half that distance, 800 pc. Based on these arguments the
distance is $600 \pm 200$~pc, and the linear size of the PL is $18 \pm
6$~pc. The magnetic field must be very smooth over this region.

If the magnetic field is also enhanced within the PL, then its
distance could be substantially smaller. For example, doubling the
field from $B_{\|} = 2~\mu$G to $4~\mu$G would lower the distance limit by
a factor of about four; the effect of changing the assumed field is
shown in the lower panel of Figure~\ref{em}. A lower distance would ease
the problem of maintaining field uniformity across the PL. 

A doubling of the field inside the PL does not seem unreasonable,
given the enhancement of electron density, which exceeds the
surroundings by at least a factor of 10. Under these circumstances the
distance would be $150 \pm 50$~pc.  While this nearer distance would
seem a reasonable choice, we caution that the PL cannot be at a very
nearby distance. If we assume that the source of the ionization is a
star of some kind, then it would probably have been detected by
existing observations. We therefore consider that the PL is at least
at the distance of the open cluster M39, which is 265 pc (see
{\S}4.4).

Striking a balance between these arguments, we place the PL at a
distance of $350 \pm 50$~pc, in a magnetic field of $B_{\|} \approx
3~\mu$G.  The excess electron density within the PL relative to its
surroundings is ${1.5}<{n_e}<{2}$~cm$^{-3}$, the emission measure is
${25}<{E}<{34}$~cm$^{-6}$pc, the radio brightness temperature is
${T_b}\approx{0.05}$~K.
At 350 pc the mass of ionized gas in the PL is 23 M$_{\odot}$.

\subsection{Are there many such objects?}

Table 1 compares the PL, which we now designate as G91.8$-$2.5, with
G137.6+1.1. The objects are similar in electron density and structure.
We note that the Faraday rotation per unit depth through the two
objects is very similar. The distances are not particularly well
determined in either case, but G91.8$-$2.5 is apparently closer and
smaller.

The similarity of G91.8$-$2.5 and G137.6+1.1 suggests that they are
two representatives of a class of object that might be quite common in
the ISM. We incline to this view, and will use the term PL generically
to refer to them. We suggest that more PLs will be detected as more of
the Galactic plane is mapped with high-resolution polarimetry.  The
mass of ionized gas required to generate a PL is from a few tens to a
few hundred M$_{\odot}$.  The ionizing radiation required to sustain
such a region can be provided by an early B star.  However,
in the two PLs discovered to date, no such star has been detected.
Furthermore, the turbulence in an \ion{H}{2} region around such a star
would probably dictate against smooth structure, and it is likely that
PLs are relatively old structures. The source of ionization for these
regions remains to be clarified, possibly through the discovery of
further examples.  Their distribution in Galactic latitude and
longitude may give clues to their origin.

\subsection{How do we detect objects of this type?}

G137.6+1.1 is detectable because it is superimposed on the \ion{H}{2}
region W5. The dense, turbulent ionized gas in W5 completely
eliminates any polarized emission from behind the \ion{H}{2} region
through its strong beam depolarization. How can we see G91.8$-$2.5
without a strong \ion{H}{2} region behind it? In this case, a strong
polarization-horizon effect in the local arm serves the equivalent
function.  At 1420 MHz, these backgrounds may be optically thin, but
they are ``Faraday thick''.

The regular polarization structure of PLs like G91.8$-$2.5 and
G137.6+1.1 can be easily obscured by the chaotic small-scale structure
that is seen in many directions. This will happen particularly
strongly if the small-scale structure is in front of the PL, but
small-scale structure behind it will also have an effect. The
detection of PLs therefore requires more than a Faraday-thick
background, it also requires a path where the structure of the MIM is
relatively unperturbed.  G137.6+1.1 appears to lie in the region
between the local and Perseus spiral arms. Magnetic fields in the
inter-arm regions can be expected to be less tangled by the disruptive
effects of star-formation. G91.8$-$2.5 is more difficult to understand.
Given the very complex structure expected within the local arm, it is
surprising that magnetic field structure along the line of sight is
sufficiently unperturbed to allow the structure to be seen. The closer
the object is to us, the easier it is to explain, and this reasoning
possibly favors the closer distance to G91.8$-$2.5.

In conclusion, PLs may be quite common in the MIM, but rather special
conditions are required for their detection.  The detection of the PL
phenomenon requires that background synchrotron emission be thoroughly
depolarized. Large \ion{H}{2} regions will do this effectively, and
the polarization-horizon phenomenon can also be effective, especially
in directions near ${l}={90^{\circ}}$ where the line of sight is
parallel to the regular field component. The PL arises from the
Faraday rotation in a smooth foreground feature of synchrotron
emission generated relatively nearby. In addition, MIM structure
along the line of sight must be relatively uniform. Detection of
nearby PLs is also easier. Only further observations will show how
often these conditions are met.

We thank Andrew Gray and Russ Taylor for their comments on a draft of
this paper.  The Dominion Radio Astrophysical Observatory is operated
as a national Facility by the National Research Council of Canada.
The Canadian Galactic Plane Survey is a Canadian project with
international partners and is supported by a grant from the Natural
Sciences and Engineering Research Council of Canada. This research has
made use of the Simbad database.

\clearpage
\begin{table}
\caption{Comparison of G91.8$-$2.5 and G137.6+1.1
\label{tbl-1}}
\begin{tabular}{lcc}
\tableline\tableline
         & G91.8$-$2.5 & G137.6+1.1 \\
\tableline
${\Delta}{\theta}~(^{\circ})$ & 100 & 280   \\
${\Delta}$RM~(rad~m$^{-2}$)   &  40 & 110   \\
Distance~(pc)                 & 350 & 1000  \\
Size~(pc)                     &  10 &   36  \\
${\Delta}n_{e}$~(cm$^{-3}$)   & 1.7 &  1.1  \\
$T_b$~(K)                    & 0.05 &  0.075  \\
E~(pc~cm$^{-6}$)              &  29 &   44  \\
Mass~(M$_{\odot}$)            &  23 &  400  \\
\tableline
\end{tabular}
\end{table}

\begin{figure}\centering
\caption{{\it Top:} Total-intensity $I$ image from the
CGPS database. {\it Bottom:} Same region in polarized intensity, $PI$.
\label{ipimap}
} 
\end{figure}

\begin{figure}\centering
\caption{{\it Top:} Stokes  $Q$ image from the
CGPS database. {\it Bottom:} Same region in Stokes $U$.
\label{qumap}
} 
\end{figure}

\begin{figure}\centering
\caption{
Images of total-intensity $I$ ({\it{top}}), polarized intensity $PI$
({\it{middle}}), and polarization angle $\theta$ ({\it{bottom}})
 towards the detected polarization structure at
$5\arcmin$ resolution. Vectors superimposed on the middle image
have length proportional to $PI$ and orientation $\theta$.
The concentric ellipses with major axes at right angles to each
other are centered at $l=91\fdg75$, $b=-2\fdg45$. 
In the upper image, the filled star symbols show the members of M39,
while outline stars show the B0 and B3 (higher $b$) stars mentioned in
the text.
}
\label{tp}
\end{figure}

\begin{figure}\centering
\caption{Variation of polarization angle along the major axis of the
larger ellipse in Figure~\ref{tp}. Angle has been ``unwrapped'' to
remove the $180^{\circ}$ ambiguity over the regions more than about
$15\arcmin$ from the center of the plot.}
\label{prof}
\end{figure}

\begin{figure}\centering
\caption{Variation of brightness temperature ({\it{top}}) and emission
measure ({\it{middle}}) as a function of the distance of a spherical
enhancement of electron density which causes a Faraday rotation
${\Delta}{\theta}$ in a magnetic field ${B_{\|}}={2~\mu}$G. The lowest
curve corresponds to ${{\Delta}{\theta}}={20^{\circ}}$ and the increment in
angle is $20^{\circ}$. The horizontal lines in the top plot indicate
the $2\sigma$ and $4\sigma$ sensitivity levels of the telescope.
The {\it{bottom}} panel shows the effect of changing magnetic field.
${{\Delta}{\theta}}={100^{\circ}}$ and all other assumptions are unchanged.
The highest curve corresponds to ${B_{\|}}={1~\mu}$G and the field increases
in steps of ${0.5~\mu}$G.
}
\label{em}
\end{figure}      

\end{document}